\documentclass[a4paper,11pt]{article}
\usepackage{pos}

\usepackage[symbol]{footmisc}
\usepackage[english]{babel}
\usepackage{graphicx}
\usepackage{graphics}
\usepackage{braket}
\usepackage{bbold}
\usepackage{amsmath}
\usepackage{nicefrac}
\usepackage{dcolumn}
\usepackage{bm}
\usepackage{slashed}
\usepackage{datetime}
\usepackage{mciteplus}
\usepackage{multirow}
\usepackage{bbold}
\usepackage{siunitx}
\usepackage{booktabs}
\usepackage{color, soul}
\usepackage[usenames,dvipsnames]{xcolor}
\usepackage{float}
\usepackage[utf8]{inputenc}
\usepackage[normalem]{ulem}
\usepackage{mathtools}
\usepackage{setspace}

\renewcommand{\thefootnote}{\fnsymbol{footnote}}

\newcommand{\tref}[1]{~\ref{#1}}

\title{Spectator-model studies for spin-dependent gluon TMD PDFs at the LHC and EIC}
\ShortTitle{Spin-dependent gluon TMD PDFs at the LHC and EIC}

\author[a,b]{Alessandro Bacchetta}
\author*[c]{Francesco Giovanni Celiberto}
\author[b]{Marco Radici}

\affiliation[a]{Dipartimento di Fisica, Universit\`a di Pavia, via Bassi 6, I-27100 Pavia}
\affiliation[b]{INFN Sezione di Pavia, via Bassi 6, I-27100 Pavia, Italy}
\affiliation[c]{Universidad de Alcalá (UAH), Departamento de Física y Matemáticas, Campus Universitario, Alcalá de Henares, E-28805, Madrid, Spain}

\emailAdd{alessandro.bacchetta@unipv.it}
\emailAdd{francesco.celiberto@uah.es}
\emailAdd{marco.radici@pv.infn.it}

\abstract{We present novel analyses on accessing the 3D gluon content of the proton via spin-dependent TMD gluon densities, calculated through the spectator-model approach. Our formalism embodies a fit-based spectator-mass modulation function, suited to catch longitudinal-momentum effects in a wide kinematic range. Particular attention is paid to the time-reversal even Boer--Mulders and the time-reversal odd Sivers functions, whose accurate knowledge, needed to perform precise 3D analyses of nucleons, motivates synergies between LHC and EIC Communities.}

\FullConference{The European Physical Society Conference on High Energy Physics (EPS-HEP2023)\\
 21-25 August 2023\\
Hamburg, Germany\\}

\begin{document}
\maketitle

\section{Opening remarks}
\label{sec:introduction}

Unraveling the inner structure of nucleons through a multi-dimensional analysis of their constituents is a frontier research of phenomenological studies at new-generation colliding machines.
The well-established collinear factorization, relying upon a one-dimensional vision of the proton content by means of collinear parton distributions functions (PDFs), has collected a long series of achievements in describing data at hadron-hadron and lepton-hadron colliders.
On the other side, fundamental problems of the dynamics of strong interactions still wait to be answered.
Understanding the origin of mass and spin of hadrons requires a stretch of our viewpoint from the collinear framework to a 3D tomographic treatment, genuinely provided by the transverse-momentum-dependent (TMD) formalism.
Our knowledge of the gluon-TMD sector is much more limited with respect to the quark-TMD case.
In~\cite{Mulders:2000sh} and then in~\cite{Meissner:2007rx,Lorce:2013pza,Boer:2016xqr} unpolarized and polarized TMD PDFs were listed for the first time, whereas first exploratory studies on gluon-TMD phenomenology were proposed in~\cite{Lu:2016vqu,Lansberg:2017dzg,Gutierrez-Reyes:2019rug,Scarpa:2019fol,COMPASS:2017ezz,DAlesio:2017rzj,DAlesio:2018rnv,DAlesio:2019qpk}.
A striking difference between collinear and TMD densities is the dependence on the gauge link.
The sensitivity of TMD PDFs to the transverse components of the gauge link makes them dependent on the process considered, while it does not happen in the collinear case~~\cite{Brodsky:2002cx,Collins:2002kn,Ji:2002aa}.
In particular, quark TMD PDFs depend on the process by means of two main gauge links: the $[+]$ \emph{staple} link, which is related to the future-pointing direction of the Wilson lines, and the $[-]$ one, which is connected to the the past-pointing one.
Conversely, the gauge-link dependence of gluon TMD is much more intricate. This leads to a more diversified kind of \emph{modified universality}.
There exist two principle gluon gauge links: the $f\text{-type}$ and the $d\text{-type}$ structure, also known in the context of small-$x$ studies as Weisz\"acker--Williams and dipole types.
The antisymmetric QCD color structure, $f_{abc}$, is typical of the $f$-type T-odd gluon-TMD correlator, whereas the $d$-type T-odd one embodies the symmetric color structure, $d_{abc}$.
The $f$-type ($d$-type) gluon TMD PDFs are sensitive to the $[\pm,\pm]$ ($[\pm,\mp]$) gauge-link combination.
%
%
A relation between the unpolarized gluon TMD function and the BFKL unintegrated distribution~\cite{Dominguez:2011wm,Hentschinski:2021lsh,Celiberto:2019slj,Nefedov:2021vvy} exists only at small-$x$ and moderate- transverse momentum~(for recent phenomenological applications see~\cite{Bolognino:2018rhb,Bolognino:2019pba,Motyka:2014lya,Brzeminski:2016lwh,Celiberto:2018muu,Bautista:2016xnp,Garcia:2019tne,Hentschinski:2020yfm,Luszczak:2022fkf,Cisek:2022yjj,Celiberto:2020wpk,Celiberto:2022grc,Celiberto:2016vhn,Bolognino:2018oth,Bolognino:2019yqj,Bolognino:2019yls,Celiberto:2020tmb,Celiberto:2020rxb,Bolognino:2021mrc,Celiberto:2021dzy,Celiberto:2021fdp,Celiberto:2022dyf,Celiberto:2015yba,Celiberto:2022gji,Celiberto:2022keu,Celiberto:2022kxx,Celiberto:2023fzz,Celiberto:2023rzw}).
Pioneering studies on quark TMD PDFs were done in~\cite{Bacchetta:2008af,Bacchetta:2010si} in the context of the spectator-model formalism. 
There, different spin states of di-quark spectators as well as several form factors for the nucleon-parton-spectator vertex were considered.
That work was recently extended to leading-twist T-even gluon TMD densities in the proton~\cite{Bacchetta:2020vty} (see also~\cite{Bacchetta:2021oht,Celiberto:2021zww,Bacchetta:2021lvw,Bacchetta:2021twk,Celiberto:2022fam}).
We present results for the T-even gluon Boer--Mulders function and a preliminary analysis on the $f$-type T-odd gluon Sivers TMD PDF, both of them obtained within the spectator-model formalism.
The Boer--Mulders TMD has particular relevance at the LHC, since it is connected to the density of linearly polarized gluons inside an unpolarized proton.
Conversely, the Sivers function is the key ingredient to study transverse single-spin asymmetries accessible at the future EIC as well as at the planned LHCspin~\cite{Aidala:2019pit,Santimaria:2021uel,Passalacqua:2022jia}.

\section{Spin-dependent gluon TMD PDFs in a spectator framework}
\label{sec:TMDs}

We model the gluon correlator according to the spectator framework. From a parent nucleon of four-momentum $P$ and mass $M$ a gluon with four-momentum $p$, transverse momentum $\boldsymbol{p}_T$, and longitudinal fraction $x$, is struck. The remainders are portrayed as an effective on-shell particle, the \emph{spectator}, with mass $M_X$ and spin-1/2.
The nucleon-gluon-spectator effective vertex embodies two form factors, given as dipolar functions of $\boldsymbol{p}_T^2$. Choosing dipolar form factors allows us to quench gluon-propagator singularities and dampen logarithmic divergences arising in the $|\boldsymbol{p}_T|$-integrated correlator.
All the spectator-model T-even gluon TMD PDFs at leading twist in the proton were calculated in~\cite{Bacchetta:2020vty}. 
In that work the pure spectator approach was improved by weighing the spectator $M_X$ mass over a continuous range via a spectral function suited to collect both the small- and the moderate-$x$ dynamics.
Model parameters encoded in the spectral mass and the nucleon-gluon-spectator coupling were determined by making a simultaneous fit of the $|\boldsymbol{p}_T|$-integrated unpolarized and helicity gluon TMD PDFs, $f_1^g$ and $g_1^g$, to the corresponding collinear densities provided by the {\tt NNPDF} collaboration~\cite{Ball:2017otu,Nocera:2014gqa} at the initial scale of $Q_0 = 1.64$ GeV. The impact of the statistical uncertainty was assessed through the well-established replica method~\cite{Forte:2002fg}.
\begin{figure}[b]
\centering

\includegraphics[width=0.4750\textwidth]{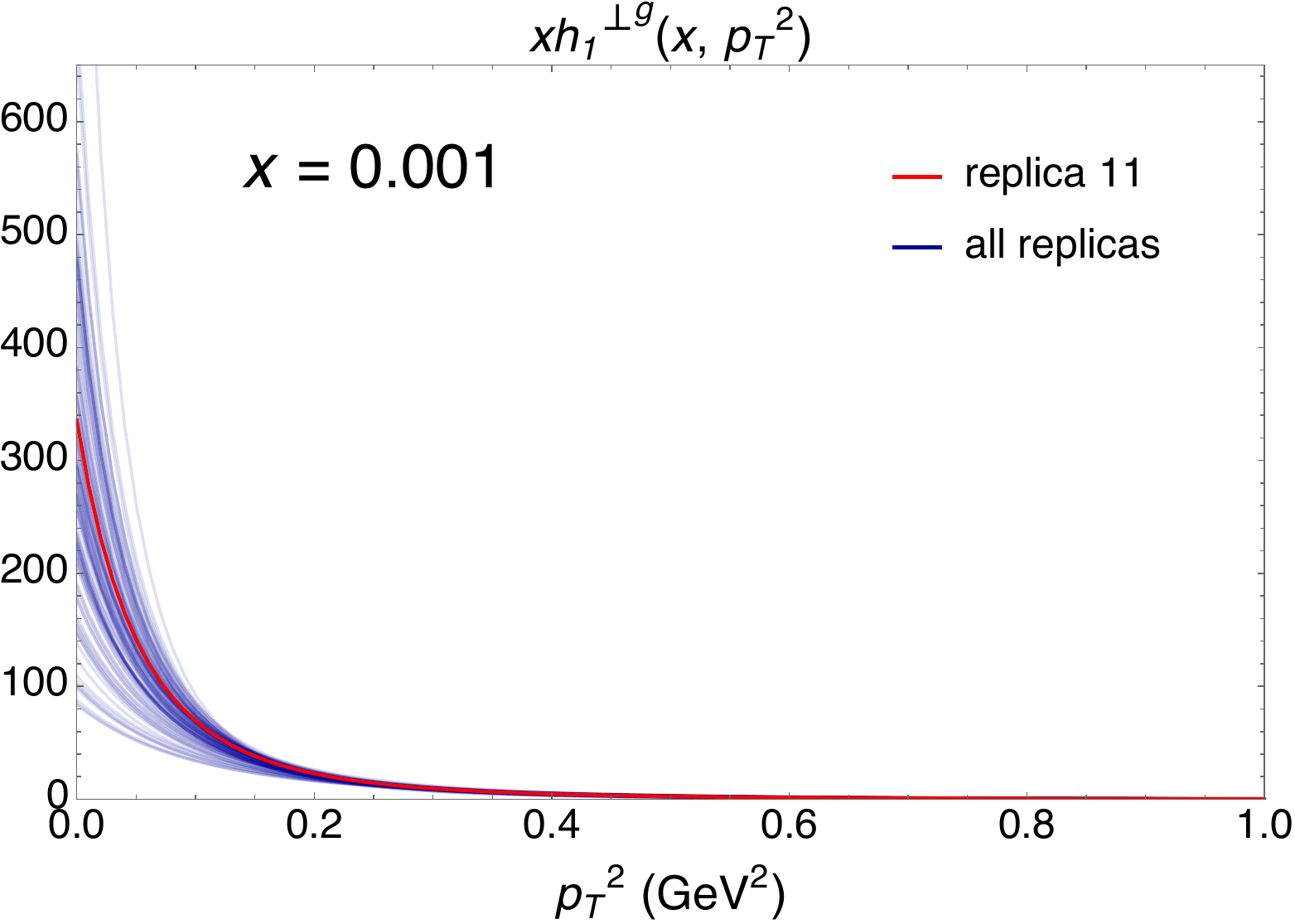} \hspace{0.40cm}
\includegraphics[width=0.4875\textwidth]{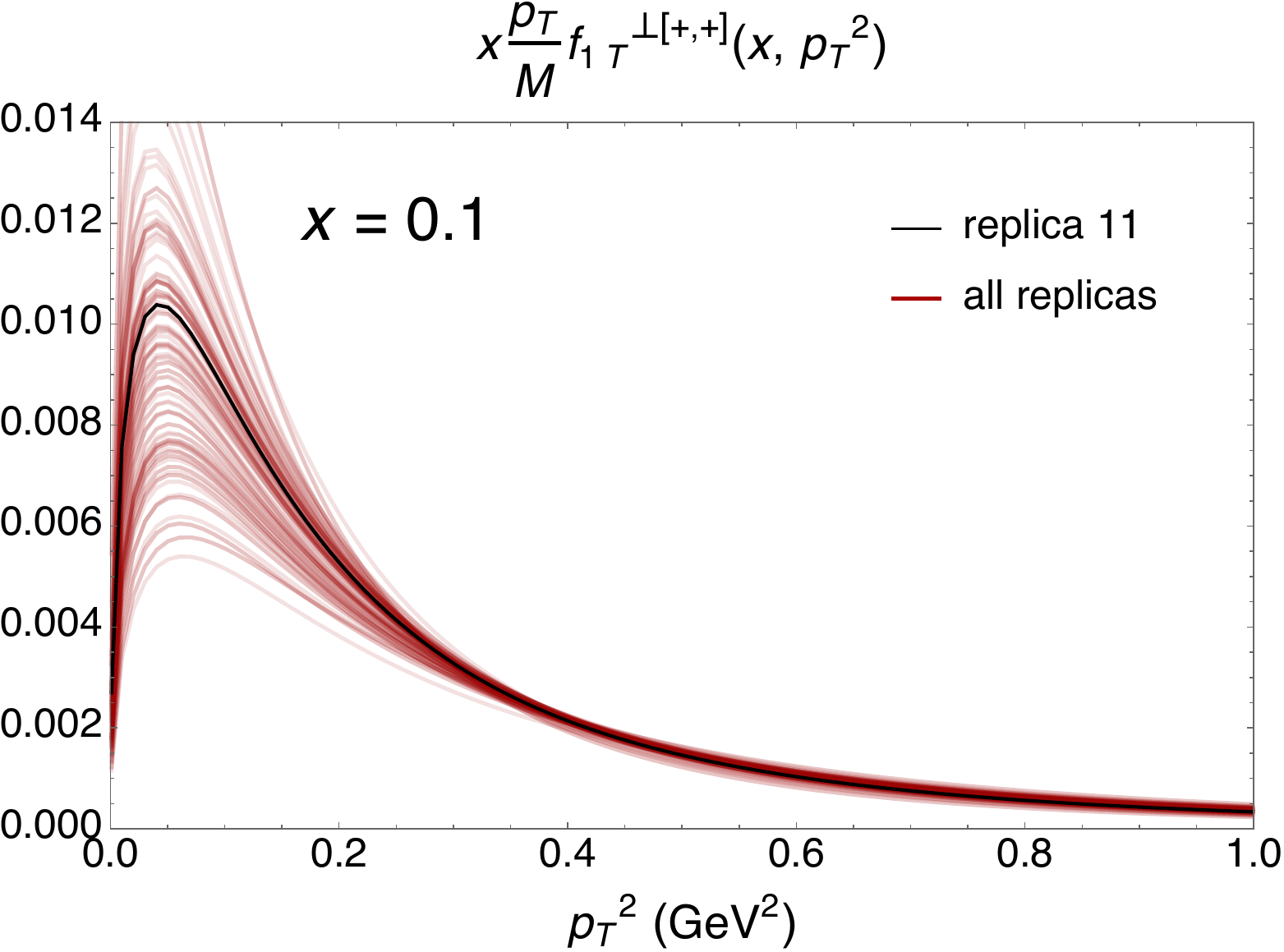}

\caption{Transverse-momentum dependence of Boer--Mulders function at $x=10^{-3}$ (left) and the $|\boldsymbol{p}_T|$-weighted $[+,+]$ Sivers function at $x=10^{-1}$ (right), obtained in the spectator model at the initial scale of $Q_0 = 1.64$ GeV. Black curve is for the most representative replica \#11.}

\label{fig:TMDs}
\end{figure}
The gluon correlator at tree level is not sensitive to the gauge link. Thus, our T-even TMD PDFs are process independent.
A T-odd density can be generated by going beyond the tree level for the gluon correlator, and taking into account its interference with another channel. Making use of the same strategy adopted in the quark-TMD case~\cite{Bacchetta:2008af}, we have considered the one-gluon exchange in the \emph{eikonal} approximation, which actually represents the truncation at the first order of the full gauge-link operator. As a consequence, the resulting T-odd structures are now sensitive to the gauge link. Therefore, they are process dependent. 
While, under gauge-link variation, a given spectator-model T-even density, say the Boer--Mulders TMD PDF ($h_1^{\perp g}$), remains the same, two distinct spectator-model T-odd functions, say two Sivers TMD PDFs ($f_{1T}^{\perp g}$), exist. 
They are obtained after suitably projecting the transverse component of the TMD gluon correlator. 
Thus one has the following relations of \emph{modified universality} for the T-odd case
\begin{align}
 \label{eq:T-eodd_TMD_PDFs}
 f_{1T}^{\perp \, [-,-]}(x, \boldsymbol{p}_T^2) \; & = \; - \, f_{1T}^{\perp \, [+,+]}(x, \boldsymbol{p}_T^2) \; ,
 \\ \nonumber
 f_{1T}^{\perp \, [-,+]}(x, \boldsymbol{p}_T^2) \; & = \; - \, f_{1T}^{\perp \, [+,-]}(x, \boldsymbol{p}_T^2) \; .
\end{align}

We present results for the spectator-model gluon Boer--Mulders function, as well as preliminary studies on the gluon Sivers TMD PDF. For the sake of consistency, values of model parameters are the ones obtained  via the fit to {\tt NNPDF} parameterizations for $f_1^g$ and $g_1^g$ integrated densities.
Plots of Fig.\tref{fig:TMDs} exhibit the transverse-momentum dependence of the Boer--Mulders function (left) at $x = 10^{-3}$ and the $|\boldsymbol{p}_T|$-weighted $[+,+]$ Sivers distribution (right) at $x = 10^{-1}$, and with the initial scale set to $Q_0 = 1.64$~GeV.
Both TMD PDFs clearly show a non-Gaussian $\boldsymbol{p}_T^2$-behavior, with a decreasing tail at high transverse momentum.
The Boer--Mulders function starts from a finite value at $\boldsymbol{p}_T^2 = 0$ and goes down very fast as $\boldsymbol{p}_T^2$ increases.
The Sivers function starts from a small, non-zero value at $\boldsymbol{p}_T^2 = 0$, then has a peak in the $\boldsymbol{p}_T^2 \lesssim 0.1$~GeV$^2$ and then a larger flattening tail.

\section{Closing statements}
\label{sec:conclusions}

We reported progress on spin-dependent gluon TMD PDFs at leading twist via an enhanced version of the spectator-model formalism, which permitted us to catch both the small- and the moderate-$x$ dynamics. We plan to complete the calculation of the T-odd gluon TMD PDFs soon. 
These studies can serve a useful guidance to access the gluon-TMD dynamics in the proton at new-generation colliding machines, such as the EIC~\cite{AbdulKhalek:2021gbh,Khalek:2022bzd,Abir:2023fpo,Bolognino:2021niq}, NICA-SPD~\cite{Arbuzov:2020cqg}, the HL-LHC~\cite{Chapon:2020heu,Amoroso:2022eow,Begel:2022kwp,Bacchetta:2022nyv} and its extension to polarized fixed targets~\cite{Lansberg:2012kf,Lansberg:2015lva,Kikola:2017hnp,Aidala:2019pit,Santimaria:2021uel,Passalacqua:2022jia}, the \emph{Forward Physics Facility} (FPF)~\cite{Anchordoqui:2021ghd,Feng:2022inv,Celiberto:2022rfj,Celiberto:2022zdg}.

\section*{Acknowledgments}
\label{sec:acknowledgments}

This work was supported by the Atracci\'on de Talento Grant n. 2022-T1/TIC-24176 of the Comunidad Aut\'onoma de Madrid, Spain.

\begingroup
\setstretch{0.9}
\bibliographystyle{bibstyle}
\bibliography{biblography}
\endgroup

\end{document}